# Observations of small-scale energetic events in the solar transition region：explosive events, UV bursts and network jets

**Zhenghua Huang, Bo Li, and Lidong Xia**

Shandong Provincial Key Laboratory of Optical Astronomy and Solar-Terrestrial Environment, Institute of Space Sciences, Shandong University, Weihai 264209, China (z.huang@sdu.edu.cn)

**Abstract:** In this paper, we review observational aspects of three common small-scale energetic events in the solar transition region (TR), namely: TR explosive events, ultraviolet bursts and jets. These events are defined in either (both) spectral or (and) imaging data. The development of multiple instruments capable of observing the TR has allowed researchers to gain numerous insights into these phenomena in recent years. These events have provided a proxy to study how mass and energy are transported between the solar chromosphere and the corona. As the physical mechanisms responsible for these small-scale events might be similar to the mechanisms responsible for large-scale phenomena, such as flares and coronal mass ejections, analysis of these events could also help our understanding of the solar atmosphere from small to large scales. The observations of these small-scale energetic events demonstrate that the TR is extremely dynamic and is a crucial layer in the solar atmosphere between the chromosphere and the corona.

## 1. The transition region of the solar atmosphere

The solar transition region (TR) is a region between the solar chromosphere and corona, which is assumed to be a thin layer (a few hundred kilometers) in the one-dimensional static models of the solar atmosphere (e.g. Vernazza et al. 1981; Avrett& Loeser 2008). Across this region, the temperature of the solar atmosphere rises from about $2\times10^4$ K in the upper chromosphere to about $10^6$ K in the corona (Mariska 1992). As a critical region between the chromosphere and corona, the TR holds the key to understanding how energy and mass are transporting from the lower to higher solar atmosphere. Therefore, it attracts a lot attention from the community. Recent observational and theoretical studies have revealed that the TR is highly dynamic and nonuniform. Figure 1 displays schematic structures of the lower solar atmosphere, which displays a range of dynamics and complexity in the region. Thus, the TR is often regarded as temperature region rather than a real geometrical layer. More details (both historical and recent advance) of the TR could be found in review articles written by Mariska (1986), Mariska (1992), Tian (2017) and Young et al.

(2018).

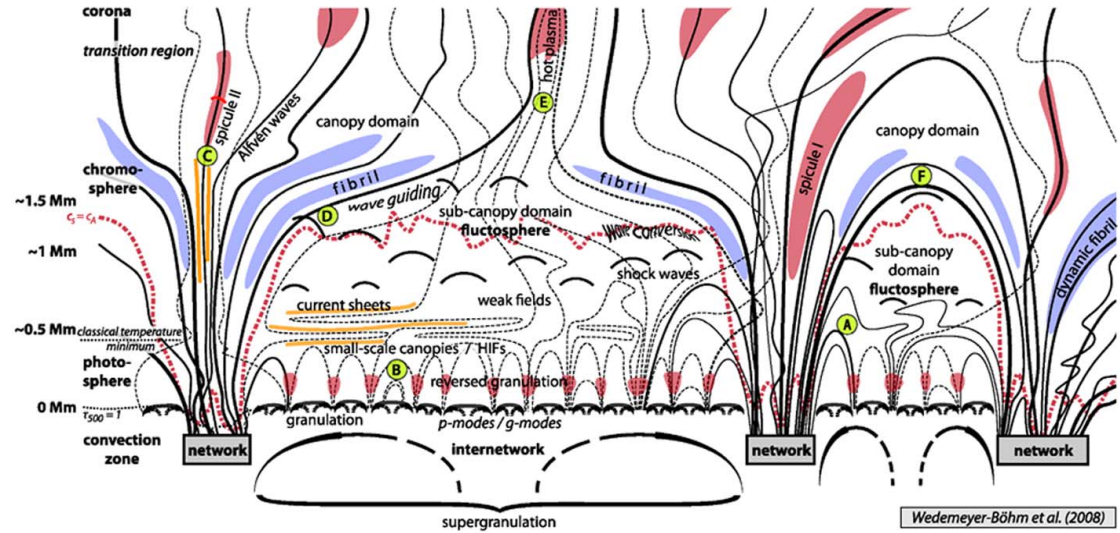

*Figure 1:* The schematic structures of the lower solar atmosphere. (Adapted from Wedemeyer-Böhm 2009).

The temperatures of the TR have determined that it mostly emits in far ultraviolet (FUV) especially by ions such as C IV, O IV and Si IV. Thus observations of the solar transition region become available only in the space era. The Naval Research Laboratory's High Resolution Telescope and Spectrograph (HRTS, Bartoe& Brueckner. 1975) recorded ultraviolet spectra of the Sun on 10 shuttle flights that revealed many dynamic phenomena in the spectral lines formed in the TR (see e.g. Brueckner & Bartoe 1983; Dere et al. 1989; 1991). Later in the 1990s, the Solar and Heliospheric Observatory (SOHO, Domingo et al. 1995) was launched carrying many instruments including two that could observe the TR, the Solar Ultraviolet Measurements of Emitted Radiation (SUMER; Wilhelm et al. 1995; Lemaire et al. 1997) and the Coronal Diagnostic Spectrometer (CDS, Harrison et al. 1995). SUMER provided full wavelength spectra from 500 Å to 1610 Å of the Sun, which have formation temperatures from $10^4$ to $10^6$ K. The SUMER provides a powerful tool for studies of the TR, and has revealed various small-scale structures of the TR in intensity, velocity, density and thermal points of view (see e.g. Hassler et al. 1999; Xia et al. 2003; 2004; Tu et al. 2005). CDS was designed to detect solar extreme ultraviolet radiation that allows us to probe conditions in the solar corona. The spectra recorded by CDS also include a few TR lines that help discover some small-scale dynamics such as blinkers (Harrison 1997) in the region. The Transition Region And Coronal Explorer (TRACE, Handy et al. 1999) launched in late 1990s provided us high spatial-temporal resolution images of the solar atmosphere that allowed us to follow the evolution of the dynamic phenomena in the TR. A few more satellites aiming at studying the Sun had been launched in the first 20 years of the $21^{st}$ century. An important mission is Hinode (Kosugi et al. 2007) launched in 2006. Hinode carried three instruments: the Extreme-ultraviolet Imaging Spectrometer (EIS, Culhane et al. 2007), the Solar Optical Telescope (SOT, Tsuneta et al. 2008) and the X-Ray Telescope (XRT, Golub et al. 2007), which could take spectra, X-ray images and magnetic field data of the Sun. These data could help understand the connection among activities seen in different temperatures from that in the photosphere to that in the corona. Another very successful mission to study the Sun is the Solar Dynamics Observatory (SDO, Pesnell et al. 2012) launched in 2010. Aboard SDO there are two instruments, the Atmospheric Imaging Assembly (AIA, Lemen et al. 2012) and the Helioseismic and Magnetic Imager (HMI, Scherrer et al. 2012), which take high spatial-temporal resolution images of the full-disc Sun. AIA takes continuous full-disc observations of the solar chromosphere and corona in seven extreme ultraviolet (EUV) channels, spanning a temperature range from approximately 20 000 to 20 million Kelvin. While HMI provides full-disc magnetic field observations of the Sun, these data help us to link different dynamics in the solar atmosphere from temperature to temperature and from location to location. Most recently, the Interface Region Imaging Spectrograph (IRIS, De Pontieu et al. 2014) was launched in 2013. It was designed to study the solar chromosphere and the TR, and it has provides fruitful observations of the TR with unprecedented spatial and spectral resolutions. IRIS has the unique capability of taking both imaging and spectral data of the TR simultaneously, and it has provided fruitful results on the dynamics of the TR (e.g. De Pontieu et al. 2014; Testa et al. 2014; Tian et al. 2014; Peter et al. 2014; Hansteen et al. 2014). The above instruments have revealed the dynamic and complex nature of the TR, which presents as many small-scale energetic events seen in imaging and spectroscopic data. In this paper, we will review on the observational facts of three types of such small-scale energetic events in the TR, namely: transition region explosive events, UV bursts and network jers.

## 2. Transition region explosive events

In the 1970s, the Naval Research Laboratory launched three flights of Black Brant sounding rockets with the High Resolution Telescope and Spectrograph (HRTS) experiment onboard (Brueckner & Bartoe 1983). Via a slit with 900 arcsecond length and 0.5 arcsec width, the experiments achieved spectral data over the wavelength range from 1175 Å to 1710 Å with a spectral resolution of 0.05 Å. The obtained spectra of Si IV ($8.0\times10^4$ K), C IV ($1.0\times10^5$ K) and O IV ($1.3\times10^5$ K) emitted from the TR show profiles with strong wings extended to 50~250 km/s toward both short and long wavelength. These events have spatial extension along the slit for 1—2 arcsec that has been taken as their spatial sizes. However, their lifetime was difficult to obtain because of the lack of series of imaging data. A recent work by Huang et al. (2014) found that the brightening corresponding to a TREE could last for about one hour. Brueckner & Bartoe (1983) proposed that such spectra are emitted from turbulent events. While Dere et al. (1984) presented their study on the TR spectral data of HRTS, they called such events explosive type of events (i.e. explosive events). The term of transition region explosive events (TREEs) starts to be widely used since Dere et al. (1989) presented a statistical analysis of the HRTS data with detailed examples of such events. In Figure 2, we present an example of TREE spectrum observed by IRIS. Although TREEs are mostly identified in the TR emissions, some of them could also produce signatures in spectra from lower temperatures, such as C I, C II, O I, Lyman series, Mg II etc. (see e.g. Dere 1992; Madjarska & Doyle 2002; Zhang et al. 2010; Huang et al. 2014b).

According to the Doppler effect, the extended strong wings in the TR spectra indicate the existence of opposite-directional plasma flows in the source. Therefore, TREEs have also been known as "bi-directional jets". Such Doppler velocities can be generated by any mechanisms that can produce bi-directional flows within the space of the pixel size. For example, swirling jets that plasma moving along magnetic field lines (Curdt and Tian 2011) and bi-directional flows along two close flux tubes (Alexander et al. 2013). Because TREEs normally are also showing enhancement in the radiation, their observations also match the picture of magnetic reconnection, which predicts a bi-directional outflows and energy release (see Priest & Forbes 2000 and Priest 2014 for theory and Su et al. 2013; Yang et al. 2015; Sun et al. 2015; Li et al. 2016; Xue et al. 2016, Huang et al. 2018a and Li et al. 2018 for real examples in the solar atmosphere). With this clue, magnetic reconnection is thought to be one of the mechanisms for producing TREEs (Dere et al. 1989, 1991; Innes et al. 1997, 2015; Li 2019), and has been evidenced by observations from instruments such as SUMER and IRIS (Innes et al. 1997; Huang et al. 2014b; Huang et al. 2018b). Figure 3 show spectral examples of TREEs as seen by SUMER raster scan. It shows that the spectra are one wing enhanced at one side, the other wing enhanced at the other side and both wing enhanced in the middle of the TREEs. These observations provide a solid evidence for the magnetic reconnection scenario as shown in Figure 4.

The magnetic reconnection scenario has been indirectly supported by observations of the phenomena associated with TREEs. TREEs usually associated with mixed polarities of magnetic features at the boundaries of the supergranulation cells (Brueckner et al. 1988; Dere et al. 1989; Dere et al. 1991; Porter & Dere 1991; Chae et al. 1998a; Madjarska & Doyle 2003; Teriaca et al. 2004; Ning et al. 2004; Muglach 2008). Moreover, Chae et al. (1998a) found that more than 60% of TREEs are associated with magnetic cancellation, a phenomenon usually occurred after magnetic reconnection (Zwaan 1987; van Ballegooijen & Martens 1989; Huang et al. 2012; Huang et al. 2014a, 2014b; Mou et al. 2016).

Moreover, many studies have focused on what dynamic features in the solar atmosphere can produce TREE spectra, and they have been found to be signatures of siphon-flows in small-scale loops (Teriaca et al. 2004), spicules (Wilhelm 2000), chromospheric upflows (Chae et al. 1998b), surges (Madjarska et al. 2009), swirling jets (Curdt & Tian 2011) and coronal brightenings and jets (Madjarska et al. 2012; Huang et al. 2012; Li et al. 2018). These observations suggest that TREEs could be triggered in ways other than magnetic reconnections.

Recently, the very high-resolution imaging and spectral data that are simultaneously taken by IRIS have revealed more details of TREEs. Huang et al. (2014b) studied a TREE that was associated with a dot-like brightening at a coronal-hole boundaries, and they found that the TREE was associated with interactions of small-scale loops and small-scale jets (see Figure 5). The authors suggest the interaction between loops could lead to magnetic reconnection that is responsible to the TREE. The connection between loop-loop interaction and TREEs has also been confirmed by a later study of Huang et al. (2015). Huang et al. (2017) make a study based on 24 TREEs found in active region aiming to find out what kinds of phenomena in the chromosphere and transition region are associated with TREEs. They found: 16 TREEs are associated with stationary loop-brightenings without any response in the chromospheric Hα spectra; six are associated with propagating loop brightenings, in which two events are associated with dark jets observed in chromospheric Hα lines; two are associated with brightenings in the conjunction region of multiple loops seen in the transition region. The study of Huang et al. (2017) clearly indicates the close relation between TREEs and the dynamics of TR loops. Since the dynamics of loops could directly link to how mass and energy are transported in the solar atmosphere (Klimchuk 2006), interaction between two loop systems might result in magnetic reconnection that could exchange plasma between the two systems and generate heats therein (Huang et al. 2018a). Such loop interactions might generate TREE spectra (see Huang et al. 2015), and thus TREE could be a signal for searching magnetic reconnection between loops in the solar atmosphere. In the other condition, a loop system might contain multiple threads that might be braided due to the random motions on the photosphere (Parker 1983a). The braiding magnetic structures could build up magnetic energy via small-scale magnetic reconnection events in the solar atmosphere (Parker 1983b, 1988; Klimchuk 2006; Schrijver 2007; Cirtain et al. 2013). In a recent work, Huang et al. (2018b) found that magnetic braidings of a transition region loop system can create plasma jets at Alfvén speed with TREE spectra found in their footpoints. The observations of Huang et al. (2018b) confirm the suggestion that magnetic braiding is also a possible mechanism to produce TREEs (Huang et al. 2017).

Based on these new observations, some modeling attempts have been put forward. Recently, Innes et al. (2015) performed a two dimensional simulation. They found that the TREE profiles can be reproduced with the multiple magnetic islands and acceleration sites that characterize the plasmoid instability if TREEs are observational signatures of magnetic reconnection sites.

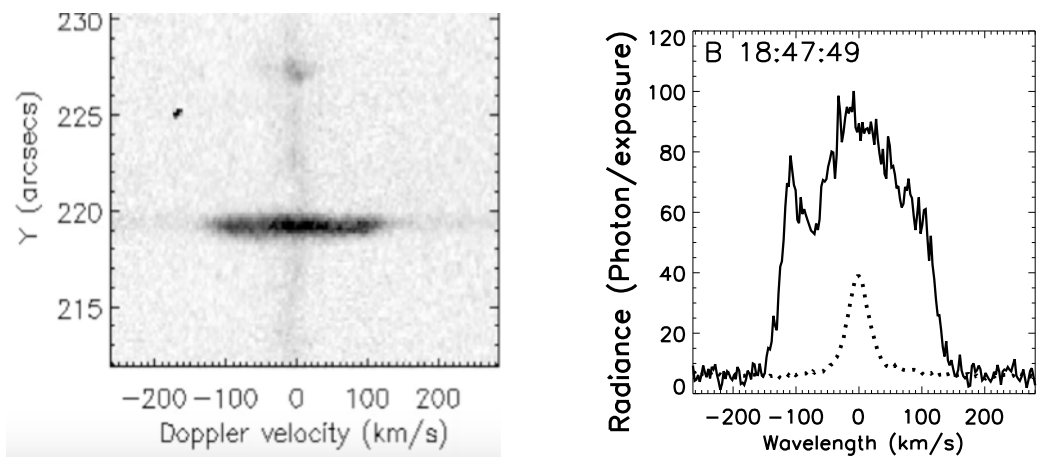

*Figure 2:* An example of Si IV spectral image of a transition region explosive event. The observations were taken by IRIS on 2013 June 27 as shown in Figure 5. Left: a slit image while it was crossing the event. The image is shown in reversed-color table, which presents strong emission in black and weak in white. The explosive event presents at the position Y=219~220 arcsec. Right: the spectrum of the event (solid line). The normal spectrum is shown in dashed line. (Adapted from Huang et al. 2014).

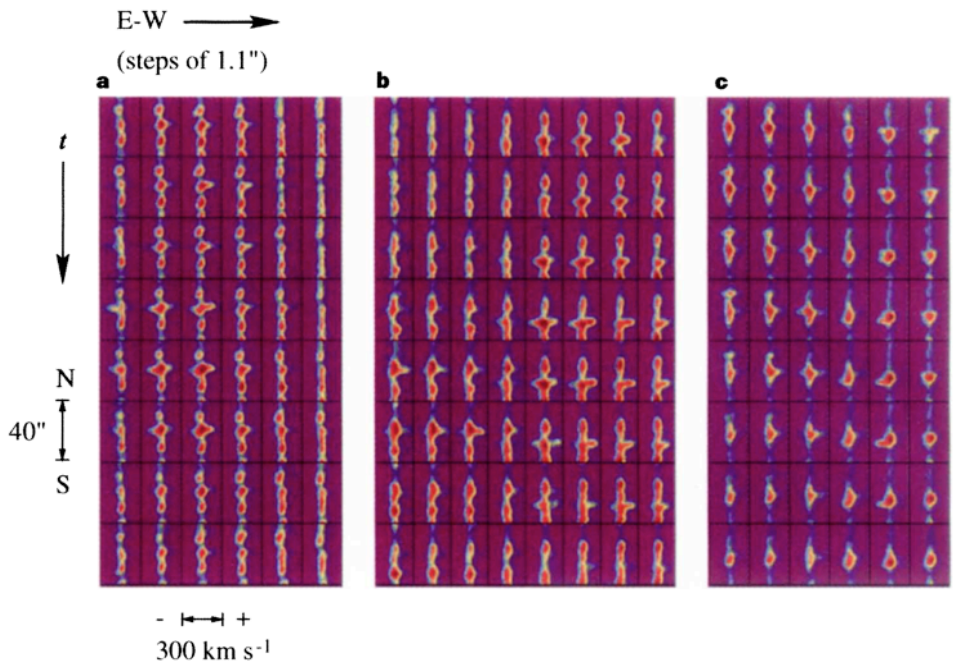

*Figure 3:* The spectra of TREEs as scanned by SUMER. (Adapted from Innes et al. 1997).

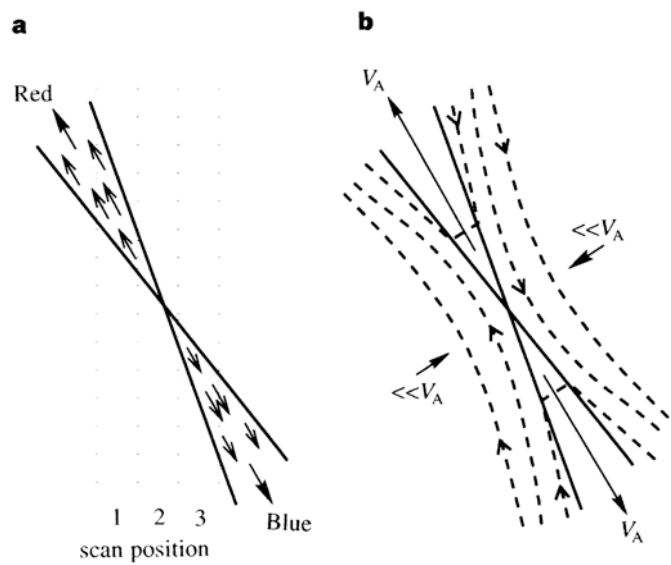

*Figure 4:* Magnetic reconnection scenario for TREEs based on the observations shown in Figure 3. While the spectrograph scanning different locations (1—3) of the reconnection site, the spectra could present blue-wing enhancement, both wing enhancement and red-wing enhancement. (Adapted from Innes et al. 1997).

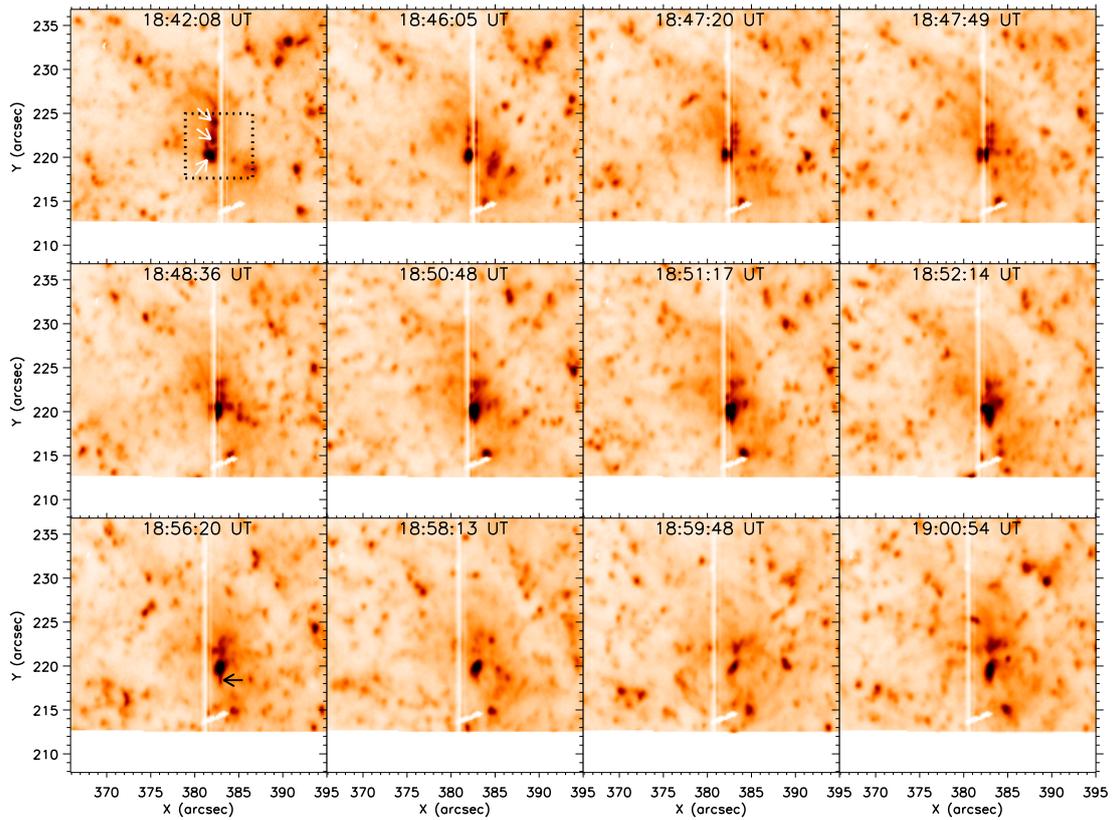

*Figure 5:* Temporal evolution of a TREE (pointed by arrows in the top-left panel) as seen in IRIS SJ 1330 Å. The images are shown in reversed color table. (Adapted from Huang et al. 2014).

## 3. UV bursts in the TR

Loop interactions in the stages of flux emergence can emit spectra with characteristics more than that of the TREEs. Peter et al. (2014) reported a new type of small-scale events seen in IRIS transition region observations, named "hot explosions" (see Figure 6 for an example). In the IRIS imaging data, these events are transient compact bright features with intensity 100—1000 times higher than the median background and a size of a few arcseconds. These events are observed in TR Si IV lines showing very broadened profiles together with clear signatures of chromospheric absorption lines (e.g. Fe II and Ni II blended in the blue wing of Si IV 1394 Å and Fe II at the red wing of Si IV 1403 Å). These events are also called "IRIS bombs" (Vissers et al. 2015; Tian et al. 2016). Since both IRIS bombs and TREEs have similar size and emit similar broad TR spectra, the blends of cold absorption lines are the main differences that distinguish these events from TREEs. An ISSI group had assembled aiming on studying these events, and they introduce a term of "UV bursts" in the TR to describe small, intense, transient brightenings in ultraviolet images of solar active regions including IRIS bombs as described above (see the definition in Young et al. 2018).

Many UV bursts are believed to be the result of magnetic reconnection in the early stage of flux emergence, when the flux tube might show a serpentine (or U-loop) geometry in the lower solar atmosphere (Fan 2001; Pariat et al. 2004, 2009; Cheung et al. 2007, 2008; Archontis et al. 2009; Cheung & Isobe 2014; Huang et al. 2018a). UV bursts in the TR are frequently observed at the early stage of flux emergence (see e.g. Peter et al. 2014; Toriumi et al. 2017; Zhao et al. 2017; Tian et al. 2018a) and might present but rare in the late phase of flux emergence (see e.g. Guglielmino et al. 2018; Huang 2018c). The UV bursts reported in Peter et al. (2014) were first suggested to be results of magnetic reconnections in the photosphere that heat plasma to TR temperature. Judge (2015) performed an independent analysis and suggested that these events are formed in low to middle chromosphere or above. While Ellerman bombs are phenomena that are well known to be signatures of magnetic reconnections in the photosphere (Ellerman 1917; Vissers et al. 2013; Reid et al. 2016; Fang et al. 2006; Hong et al. 2014; Nelson et al. 2013, 2015, 2016; Fang et al. 2017; Ni et al. 2016, 2018), Tian et al. (2016) crosscheck IRIS observations and Hα images from the Chinese New Vacuum Solar Telescope (Xu et al. 2013; Liu et al. 2014) of 10 TR UV burst events, and they found that 6 out of 10 events are associated with Ellerman bombs in the photosphere and the rest 4 are not. This indicates that some UV bursts in the TR are signatures of magnetic reconnections in the photosphere that heat plasma up to at least the TR temperature (Kim et al. 2015; Tian et al. 2016). Such a conclusion has also been confirmed by Nelson et al. 2017, who found some Ellerman bombs in quiet Sun also have responses in the IRIS Si IV lines. It is worth to mention that the blends of chromospheric absorption lines in the TR emission lines might also be the results of cold plasma (such as filaments, mini-filaments) at the foreground (Vissers et al. 2015; Huang et al. 2017; Hou et al. 2019), and thus it is not complete criteria for this species of UV bursts.

The species of TR UV bursts are also diverse. Hou et al. (2016) reported a different type of UV bursts, so-called "Narrow-line-width UV bursts" (or "NUBs" for short). NUBs are also compact brightenings but showing very different characteristics in the TR spectra. The TR spectra (Si IV) of NUBs are regularly Gaussian but extremely narrow (less than half of that from a normal region). NUBs are normally found locating above sunspots and might consist of multiple bright cores.

With these observations, some simulation experiments have tried to explain whether magnetic reconnection in the photosphere could heat plasma to transition region temperatures. Ni et al. (2018a,b) found that the initially weakly ionized plasmas can become fully ionized within the reconnection region and the current sheet can be strongly heated to above $2.5\times10^4$ K in the cases that the reconnecting magnetic field is in excess of a kilogauss and the plasma β is lower than 0.0145. For more details of UV bursts, we refer to a review paper by Young et al. (2018).

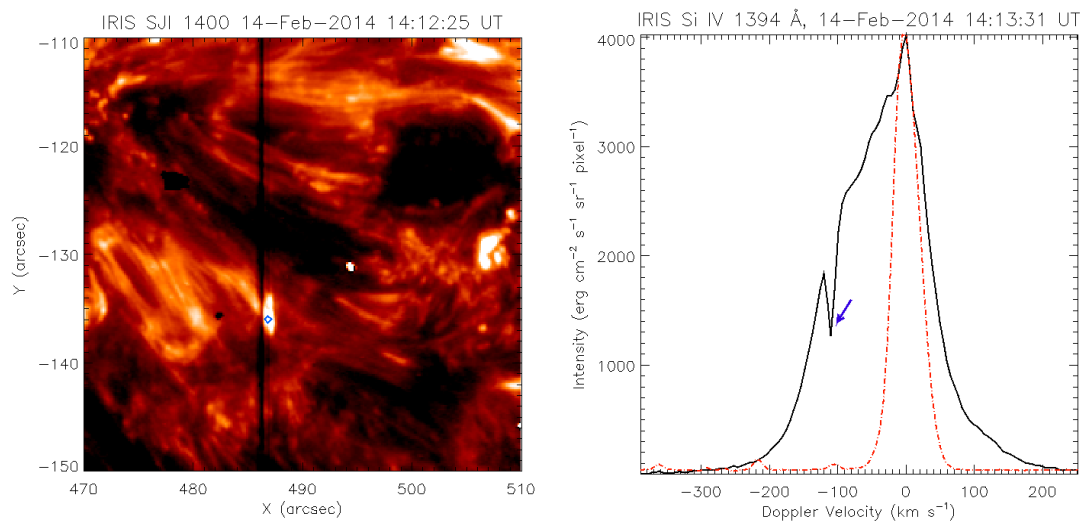

Figure 6: An example of IRIS bomb. The IRIS bomb is the compact bright feature marked by blue diamond in the IRIS SJ 1400 Å image (left panel). The IRIS Si IV 1394 Å spectrum sampled from the location of the blue diamond is shown in the right panel (black solid line). The Blue arrow points at the location where the absorption lines of Ni II is superimposed in the wing of the Si IV line. The red dashed-dotted line in the right panel is the referent spectrum that is taken from a quiet region and multiplied by 9 times. Courtesy of Mr. Zhenyong Hou.

## 4. Jets in the TR

Besides the energetic events outstanding from the brightness and spectra, jets also frequently occur in the TR. The IRIS observations have revealed that jets are ubiquitous in the TR network regions and so-called network jets (Tian et al. 2014, and see Figure 7 for examples). The network jets in coronal hole regions are small-scaled with widths of ≤300 km, lifetimes 20—80 seconds and speeds of 80—250 km/s (See Figure 8 and Tian et al. 2014; Qi et al. 2019). Network jets have also been found in quiet-sun region. These jets are typically slower and shorter than those in the coronal holes (Narang et al. 2016; Kayshap et al. 2018). Some of the network jets should be powered by magnetic reconnection and many could be heated to at least $10^5$ K (Tian et al. 2014). The appearances, sizes, lifetimes and speeds of the network jets have large overlap with that of spicules (see characteristics of spicules in e.g. Xia et al. 2005; de Pontieu et al. 2007; Zhang et al. 2012; Pereira et al. 2012; Pereira et al. 2014). Since both network jets and spicules originate in the network regions, it suggests that at least some of spicules in the chromosphere should be heated to TR temperature and shown as network jets. A recent simulation by Yang et al. (2018) based on magnetic reconnection driven by a combination of magnetic flux emergence and horizontal advection shows that both fast warm jet (much similar to the network jets) and slow cool jets (mostly like classical spicules) could be launched at the same time.

Recently, Chen et al. (2019) carried out a statistic analysis aiming on searching for the connection between TREEs and network jets. They found: (1) TREEs with double peaks or enhancements in both wings appear to be located at either the footpoints of network jets, or transient compact brightenings. These TREEs are most likely originated from magnetic reconnection. (2) TREEs with enhancements only at the blue wing are mainly located above network jets along their propagating directions. These TREEs clearly result from the superposition of the high-speed network jets on the TR background. (3) TREEs showing enhancement only at the red wing of the line are generally located around the footpoints of network jets, likely caused by the superposition of reconnection downflows on the background emission. Their analysis suggests that some TREEs are related to the birth or propagation of network jets, and that other TREEs are not connected to network jets.

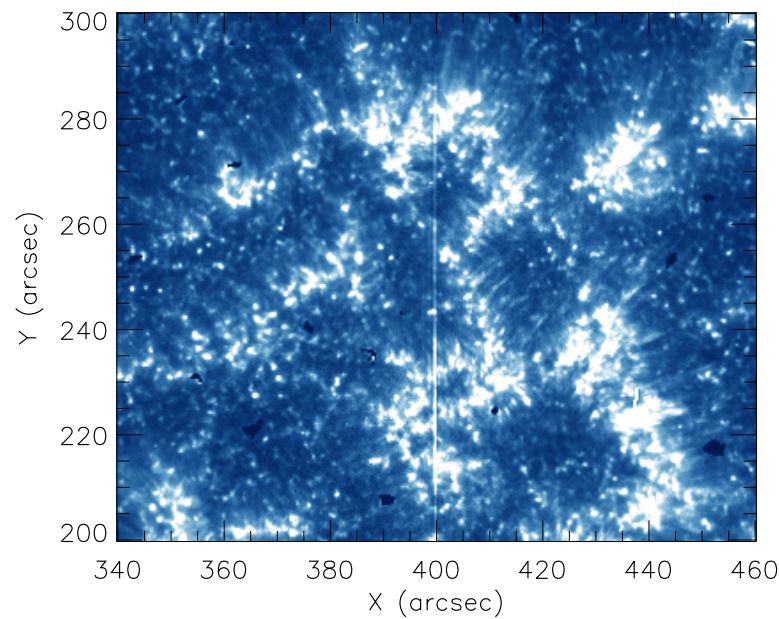

*Figure 7:* A network region observed in IRIS SJ 1330 Å passband on 2015 December 4 at 01:29 UT. The compact brightenings are organized to be the so-called "network lanes". The elongated thin features extending from the bright network lanes are network jets. Courtesy of Ms. Youqian Qi.

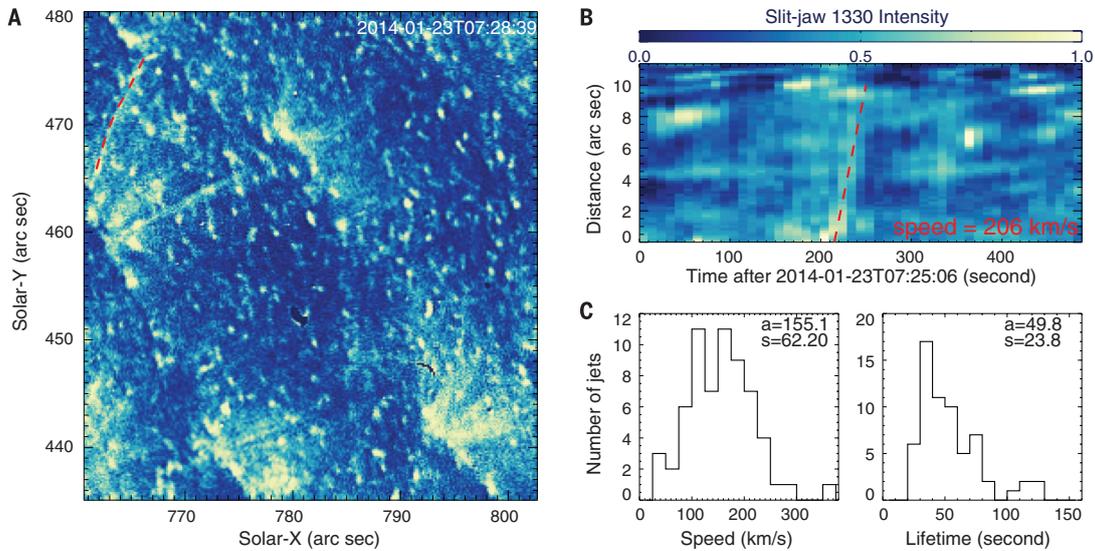

*Figure 8:* Network and jets seen in IRIS SJ 1330 Å (A), time-distance diagram of a jet (B) and statistic histograms of the speeds and lifttimes of network jets. (Adapted from Tian et al. 2014).

Furthermore, Qi et al. (2019) make a statistical study of network jets and found that network jets in the footpoints of strong coronal plumes are more dynamic (higher and faster) than those from regions without clear coronal plumes. This suggests that more dynamic network jets might somehow connect to hotter features in the corona. It has been reported that spicules (De Pontieu et al. 2011; Pant et al. 2015; Jiao et al. 2015; Samanta et al. 2015; Jiao et al. 2016) and/or shocks (Hou et al. 2018) could be the possible sources of the propagating disturbances in the coronal plumes. It is likely that some of the network jets should also be heated to coronal temperatures.

Besides network jets, there are also some other types of small-scale jets in various regions of the TR. Huang et al. (2018b) reported very fine jets occurred in an eruption of a braiding structure that involves quasi-open field. The jets have speeds more than 300 km/s that are comparable to the Alfvén speed in the eruptive region. The observations shown in Huang et al. (2018b) also demonstrate that these jets are results of magnetic reconnections due to magnetic braiding. Recently, Tian et al. (2018b) reported high-speed jets found above the light bridge of sunspots. These jets could be clearly seen in images taken at offbands of Hα and they also show signatures of bi-directional flows with speeds as large as 200 km/s in the TR spectral observations. These jets also show clear inverted-Y shapes that are clear evidence of magnetic reconnection between close magnetic field and open field.

5. Summary

The solar transition region (TR) is the crucial interface region between the chromosphere and corona. It is normally defined in the temperature regime from about $2\times10^4$ K to about $10^6$ K. The space-born instruments have opened windows for studying this region and many dynamic phenomena have been discovered. This paper reviews a few types of small-scale energetic events occurring in the TR. These events are defined in either spectral or imaging data. Multi-instrument observations available recently have given opportunities to view the same phenomenon in both spectral and imaging data, and it has given more insights to the phenomenon. The observations of these small-scale energetic events indicate the extremely dynamic and complex nature of the TR and they demonstrate that the dynamics of the TR is crucial in the processes of mass and energy transportation between the solar chromosphere and the corona.

**Acknowledgments**

The authors thank the organizing committee of the International Workshop of "Eruptive energy release processes on the Sun and stars: origins and effects" for organizing the meeting, especially Dr. Larisa Kashapova for the invitation to write this review. The authors are supported by National Natural Science Foundation of China under contracts of 11761141002, U1831112, 41627806, 41474150 and 41404135. Z.H. thanks the Young Scholar Program of Shandong University, Weihai (2017WHWLJH07). IRIS is a NASA small explorer mission developed and operated by LMSAL with mission operations executed at NASA Ames Research center and major contributions to downlink communications funded by ESA and the Norwegian Space Centre.